# Mechanical behaviour of additively manufactured Ti6Al4V meta-crystals containing multi-scale hierarchical lattice structures

*J. Lertthanasarn\*, C. Liu, MS. Pham*

Dept. of Materials, Royal School of Mines, Imperial College London, London SW7 2AZ, United Kingdom
\* Corresponding author: j.lertthanasarn17@imperial.ac.uk

Abstract

The mimicry of crystalline microstructure at meso-scale creates a new class of architected materials, termed meta-crystals, and offers effective ways to significantly improve the toughness and eliminate the post-yield collapse of architected materials. This study investigated the mechanical behaviour of polygrain-like meta-crystals fabricated from Ti6Al4V by laser powder bed fusion. The mechanical behaviour of Ti6Al4V meta-crystals is governed by lattice structures across length-scales: the crystalline microstructure, architected crystal-like mesostructures and the quality of lattice struts. Due to the intricate architecture, significant processing defects were seen in the printed meta-crystals, in particular notch-like defects due to lack of fusion at the free surface of struts. Such defects raised stress concentration and reduced the load-bearing area of struts, hence significantly weakening the lattice struts. In addition, the as-printed condition was brittle due to the presence of acicular $\alpha'$ martensites. The defects and the as-printed brittleness led to the premature fracture of struts and compromised the benefits of the crystal-inspired mesostructures. The study subsequently conducted multiple measures to resolve this issue: increasing the strut diameter to reduce the influence of the process defects and annealing to relieve internal stresses and regain ductility. The combination of the increase in strut diameter and annealing successfully enabled the crystal-like architected mesostructure to effectively improve the toughness of the meta-crystals.

1      Introduction

Architected lattice materials are a versatile type of mechanical meta-material with great potential for structural applications due to their excellent specific strength and capability of achieving unprecedented properties. Their architecture consists of a connected network of struts which can be distributed periodically or stochastically. Lattice materials are defined by their unit cell (the smallest arrangement of struts that represent the symmetry of the whole lattice structure) and their lattice parameters (such as the strut diameter and lattice spacing). The deformation mechanism can be predicted from the unit cell configuration. Typically, stretching-dominated lattices are stronger, while bending-dominated lattices offers better post-yield stability (Deshpande et al., 2001a). The mechanical behaviours of stretching-dominated lattice materials such as the octet-truss (Deshpande et al., 2001b), TPMS (Al-ketan et al., 2018), and bending-dominated lattice materials such as the BCC (Torrents et al., 2012), tetrakaidecahedral (Sullivan et al., 2008) were mostly improved by altering the lattice parameter.

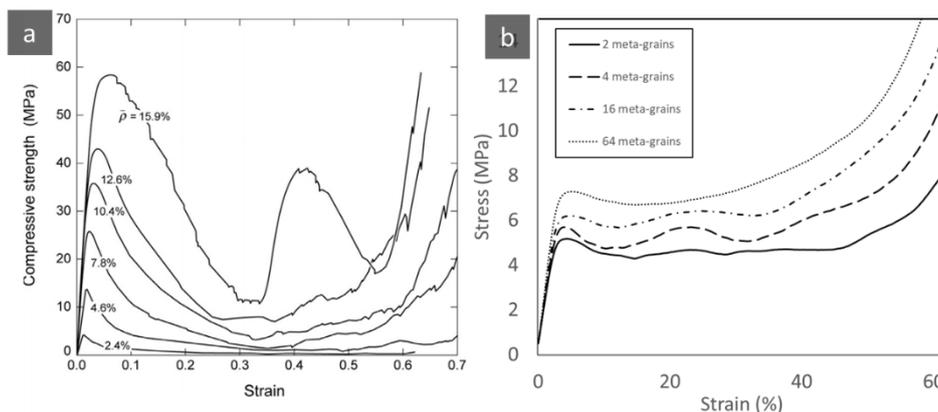

*Figure 1 - Stress-strain response of (a) Ti6Al4V octet-truss lattices with varying relative densities* (Dong et al., 2015), *reproduced with permission from Elsevier, and (b) polymer meta-crystals with increasing number of meta-grains.*

The stress-strain response in Fig 1a is typical of stretching-dominated lattice materials with varying relative densities. There are generally three distinct regions of the compressive response: (1) the initial linear elastic region followed by (2) the sudden loss in strength (with a period of mechanical instability) and lastly (3) the densification. The axial compression of struts leads to the high peak stress, but also to the mechanical instability due to failure by buckling; both are characteristics of periodic stretching-dominated lattice materials. Despite this, it was recently showed that this instability can be avoided with lattice structures that mimic the crystalline microstructure found in crystals, termed meta-crystals (Pham et al., 2019). The meta-crystals showed improved specific strength and stable post-yield behaviour. The study shows that the boundaries between polygrain-like domains were able to arrest and deflect the localised deformation bands that are the main source of the mechanical instability, thus significantly improving the post-yield stability and strength of polygrain-like architected lattice materials (Figure 1b) (Pham et al., 2019). For critical load-bearing applications such as the aerospace industry, high strength metallic alloys are often used due to their favourable mechanical properties. It is, therefore, important to study how the meta-crystal approach can enable the design of lightweight and damage-tolerant metallic meta-materials. In addition, the fabrication of polycrystal-like lattice materials by a metallic alloy results in multi-scale hierarchical lattice structures consisting the polycrystalline microstructure (including the atomic lattice) and polycrystal-like mesostructures. Such a hierarchical structure at each length scale has its own role in the mechanical behaviour of meta-crystals, providing additional degree of controlling the meta-crystal response. Therefore, in-depth understanding of the effects of the crystalline and crystal-like structures is crucial for development of lightweight, high strength, and damage tolerant meta-crystals.

Additive manufacturing (AM) has been used to fabricate lattice structures from a range of materials including polymers such as photocurable resins (Ling et al., 2019) and PLA (Niknam et al., 2018), metals such as 316L stainless steel (Gümrük and Mines, 2013) and Ti6Al4V (Li et al., 2012), as well as ceramic lattices such as $Al_2O_3$ (Zheng et al., 2014) and SiOC (Eckel et al., 2015). Powder bed fusion (PBF) is the most commonly used metal AM technique for fabrication of metallic architected materials, for example (Gümrük and Mines, 2013; Liu et al., 2017; Van Bael et al., 2011). Numerous defects are inherent to PBF processes due to violent heating, and complex melting and cooling phenomena (Sames et al., 2016). Physical defects such as pores, cracks, and the discretisation of the geometry into layers, as well as residual stresses from complex thermal cycles are detrimental to the mechanical properties of PBF-produced parts. Although defects can be minimised by optimising the processing parameters, it remains challenging to additively manufacture high quality metallic materials, especially intricate structures. Such challenges may limit the contributions from the architecture design and intrinsic microstructure of the base material. This study investigates the significance of such defects on the mechanical behaviour of Ti6Al4V polycrystal-inspired meta-crystals fabricated via Laser PBF; to what extent their presence alters the efficacy of crystal-inspired designs. The study also explored various ways to mitigate the adverse effect of process defects to enhance the contribution of crystal-like mesostructured features. Ti6Al4V was chosen to fabricate polygrain-like meta-crystals for two reasons: due to its wide use in industry such as aerospace and orthopaedic devices thanks its high strength yet low density characteristics (Leyens and Peters, 2010). As mentioned above, the use of Ti6Al4V results in meta-crystals containing multi-scale hierarchical structures from the crystalline microstructure to crystal-like architected lattice mesostructures. On the crystalline micro-scale, the microstructure of AM Ti6Al4V usually contains fine solidification microstructure, dense dislocations (Pham et al., 2020) and the *α'* martensite (Xu et al., 2015). An annealing heat-treatment of the Ti6Al4V alloy above the beta-transus temperature (995°C (Lütjering and Williams, 2007)) changes the martensitic microstructure to an (*α*+*β*) microstructure (Vrancken et al., 2012). Therefore, the heat treatment was used in this study to alter the crystalline microstructure and study the effect of different combinations of crystalline microstructure and architected mesostructures on the behaviour of Ti6Al4V meta-crystals. Finally, the study discusses how to create a synergistic measure in effectively integrating the meta-crystal approach to improve the damage tolerance of Ti6Al4V architected materials.

## 2    Methods

### 2.1    Polygrain meta-crystal design

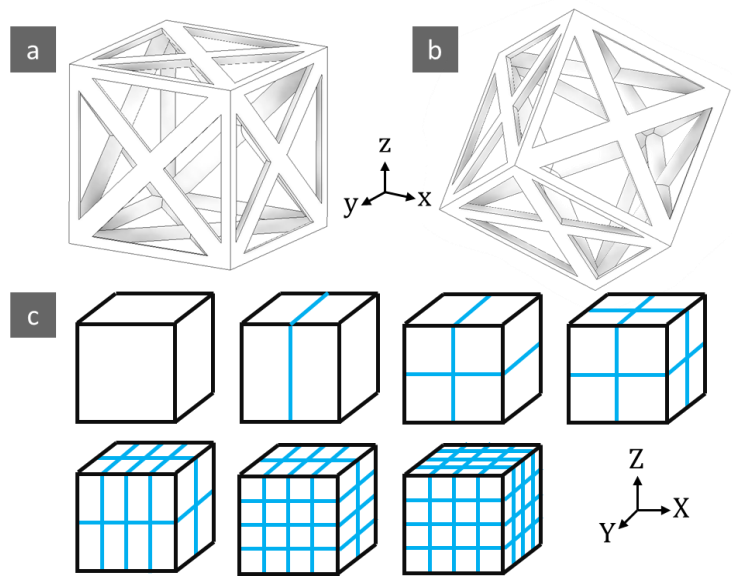

*Figure 2 - (a) FCC unit-cell (b) Rotation of unit-cell 45° about global x- and y-axis (c) Partition into domains for meta-grain and location of square boundary lattice insert*

Polycrystal-inspired architected materials (i.e. meta-crystals), were designed by the software nTopology Element. The overall morphology of all polygrain-like meta-crystals is cubic with the same nominal dimensions (40x40x40 mm); however, their volumes were divided into different numbers of cuboidal domains (called meta-grains) each mimicked a grain of a crystal and was infilled by an internal lattice structure. The unit cell of the lattice in this study was an FCC-like configuration (Fig. 2a) with an edge length (i.e. lattice spacing) of 5 mm. The unit cell was rotated ±45° about the global X- then Y-axis before being tessellated into each domain (Fig. 2b). The rotations of each meta-grain ensured the lattice structure was inclined at the same angle to the loading direction, while maintaining a misorientation between adjacent meta-grains. The square planar lattices inserted at the boundary between meta-grains provide attachment points for 'open' struts (struts connected only at one end). This study investigated meta-crystals with 1, 2, 4, 8, 16, 32, and 64 meta-grains; the arrangement of the meta-grain domains in each meta-crystal are shown in Fig. 2c.

*2.2  Fabrication and heat treatment*

The architected lattice materials were fabricated via laser powder bed fusion (L-PBF) from Ti6Al4V, courtesy of Betatype Ltd. The meta-crystals were fabricated with two different strut diameters: 0.5 and 0.8 mm. The processing parameters were consistently the same for all meta-crystals. The laser traced in a Contour-Hatch strategy using a meander hatch pattern with 95 µm spacing; the hatch direction rotated 67° between layers. The hatch volume used a 200 W modulated laser with a spot size of 75 µm, a point distance of 60 µm, and an exposure time of 70 ms. The layer thickness was 60 µm.

A set of meta-crystals were annealed to relieve internal stresses and increase ductility. Heat-treatments were performed inside a vacuum furnace to minimise oxidation. The samples were heated at a rate of 20°C/min, followed by dwelling at 1050°C above the transus temperature for 2 hrs to homogenise the microstructure, and cooled in the furnace (Silva and Gibson, 1997)(Hernández-Nava et al., 2016).

## 2.3 Defects and microstructure characterisation

Microstructure of the base material in the as-built and after heat-treatment was studied by scanning electron microscope (SEM) and electron backscatter diffraction (EBSD). Struts were sectioned and ground down incrementally using SiC paper with grit sizes from 600 to 4000. After which, the samples were alternatively polished and etched six times. Struers' MD-Chem polishing pad and neutralised OP-S (10:7:3 by vol. of H20, OP-S, H2O2) was used for the polishing. And for etching, the samples were swilled in Krolls' reagent (92:5:3 by vol. of H20, HF, HNO3) for 10 seconds. A Zeiss Gemini Sigma300 FEG SEM system equipped with a Bruker's Argus high-definition backscatter detector was used. SEM images were captured with at an accelerating voltage of 5 and 10 keV with an aperture size of 30 μm, while the EBSD maps were scanned at 20 keV with an aperture size of 120 μm and a step size of 0.5 μm.

X-Ray Computed Tomography (xCT) was performed to examine the geometry and internal porosity of meta-crystals with the 0.5 mm strut diameter. Scans were obtained at two different magnifications, giving voxel resolutions of 35 μm and 10 μm. The xCT parameters were 210 kV and 10W with 1 second exposure, 8-frame averaging and 1 mm of tin for filtration. Reconstruction of xCT data was done using ImageJ to quantify the volumes, pores, and other geometrical defects. The resolution of the xCT was 10.3 μm in all directions and, therefore, features smaller than the resolution was not represented.

## 2.4 Mechanical testing and analysis

Quasi-static compression tests were conducted in the Instron 6620 universal testing machine. Both the top and bottom platens were greased before each test to reduce friction with Castrol multipurpose high temperature grease. The meta-crystals were compressed at a constant strain rate of 0.001 s$^{-1}$ along the same global z-direction.

Deformation behaviour of meta-crystals during the compression tests were recorded in 1 second intervals using a Nikon DSLR camera. The deformation field was then analysed by digital image correlation (DIC) using the software DaVis. A subset with a size of 101 x 101 pixels was used to analyse the images sized 6000 x 4000 pixels with a step size of 25 pixels.

Stresses were calculated using the average nominal area of the meta-crystals, measured using Vernier callipers, and normalised by the measured relative density, $\bar{\rho}$. The measured $\bar{\rho}$ was a ratio of the density of a fabricated material (with its mass being weighed with a digital scale and its volume being calculated from the average nominal dimensions) to the base density that was taken from literature (Lütjering and Williams, 2007). The theoretical relative density is calculated by the volume ratio of meta-crystals to the global dimensions from the design. The initial collapse is quantified with the percentage of the first stress drop (Eqn. 1). The toughness was measured as the area under the normalised stress strain curve (i.e. energy absorbed per unit volume) up to 60% strain (which was nearly the onset of densification in this study).

$$stress\ drop\ \% = \frac{100(peak\ stress - trough\ stress)}{peak\ stress} \qquad (1)$$

3     Results

*3.1    Printing quality: Geometric accuracy and internal defects*

Four sets of polygrain-like meta-crystals with 1, 2, 4, 8 16, 32, and 64 meta-grains were fabricated in total. Each set differed by strut diameter and post-processing treatment, specifically: 0.5 mm/as-printed, 0.5 mm/annealed, 0.8 mm/as-printed, and 0.8 mm/annealed. The difference between the measured and theoretical $\bar{\rho}$ indicates the print quality of the lattice struts: the higher the difference, the lower the quality. Tomographic and microscopic observations revealed that deviations from the theoretical relative density resulted mainly from the geometric inaccuracy from the design and surface defects (Figure 4). There is negligible difference between the as-printed and the annealed meta-crystals indicating that no change in shape occurred after the heat treatment.

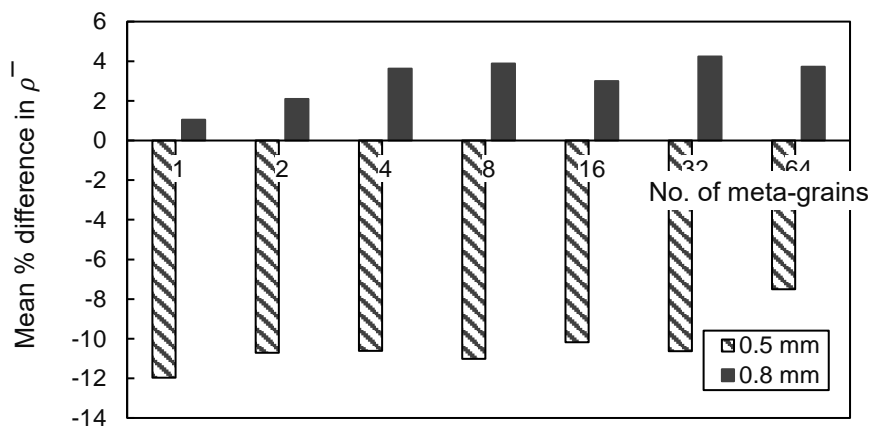

*Figure 3 - The average percentage difference in the relative density theoretically calculated and the measured value for all as-print meta-crystals with 0.5 and 0.8 mm strut diameters*

Figure 3 suggests that the print quality of the 0.8 mm diameter struts was better than the 0.5 mm diameter struts, as the absolute difference between the $\bar{\rho}$ is smaller. This was corroborated by representative micrographs of the 0.5 and 0.8 mm diameter struts in Figure 4a and b, which shows smaller geometrical deviation and fewer pores for the 0.8 mm struts. The porosity and surface roughness of the struts are clearly distinct between the two strut diameters. The quantity of internal pores is much higher for the smaller diameter struts in both the node and the strut regions. Fig. 4a and b revealed that most of the pores observed are lack of fusion pores, which was also the cause of the high surface roughness if the lack of fusion occurs at the surface of the struts. The 0.5 mm struts also showed a much more irregular surface, sharp notches on free surface and large deviations in the diameter (from the nominal dimension) along the length of the strut; such notch-like defects are stress concentrators and can significantly weaken the struts. There are no clear trends in the quality of prints with increasing numbers of meta-grains.

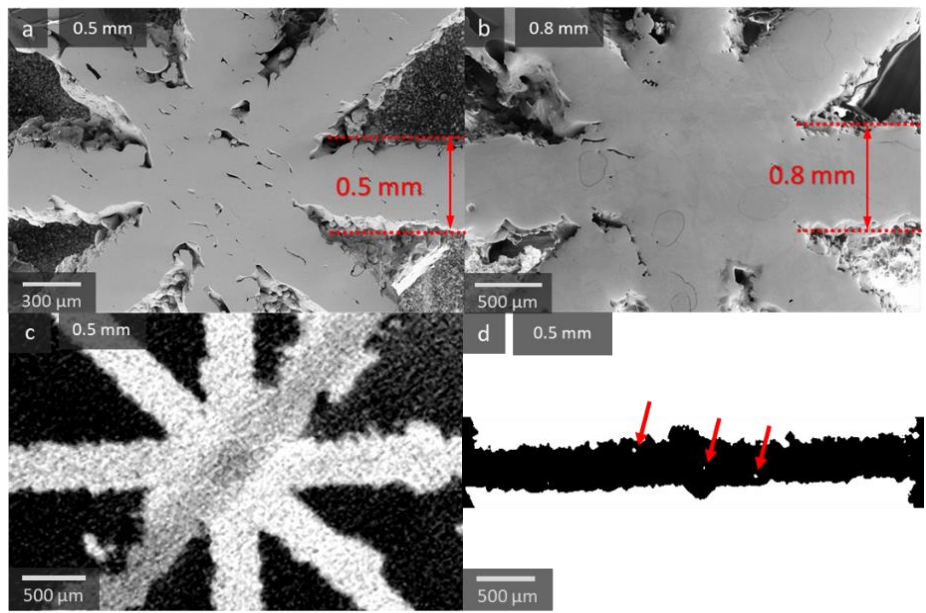

Figure 4 - SEM micrographs showing a longitudinal section of the (a) 0.5 mm and (b) 0.8 mm diameter struts with the red dotted lines depicting designed strut diameter for comparison, (c) Longitudinal sections of the 0.5 mm diameter struts obtained via xCT, (d) processed xCT data of individual strut to reveal internal pores.

xCT scans were acquired for further quantification of the defects in particular internal pores. Only scans of the meta-crystals with 0.5 mm diameter struts were acquired due to their higher density of defects. As with the previous analysis of the $\bar{\rho}$, the xCT data also revealed a similar difference between the theoretical and actual volume of the lattice, consistent with the observations shown in Figure 3. Compared to the theoretical volume from the computer models, the real volume is approximately 24% smaller. To analyse at a higher resolution and obtain more details, the xCT data of some struts were analysed in detail. The volume of the selected struts measured from the xCT data was consistently less than the theoretical volumes by 21.5% on average, approximately double the difference given by the measured $\bar{\rho}$ but slightly smaller than the overall difference determined from the xCT data of the entire structure. The latter suggests that the volume of the struts was less than the designed whereas the volume of nodes was more than the designed. This is supported by xCT and the SEM images in Figure 4 with shows enlarged nodes. The enlarged nodes are likely caused by adjacent struts overlapping at the intersection; therefore, increasing the strut diameter can lead to larger the nodes and shorter strut lengths. In addition to the volume of the struts, the volume of internal pores in the 0.5 mm strut was also calculated from the processed xCT data (Fig. 4d). The average volume fraction of the internal pores is 0.32%, showing little deviations between different meta-grain designs and strut orientations. The average minimum diameter of the struts printed in the same orientation were consistent with each other, while the average minimum ranged from 0.41 to 0.48 mm at different strut orientations. SEM images of the fracture surface confirms the severity of the reduction in diameter with failure occurring at the smallest diameter at which the consolidation was poor (Fig. 5), suggesting notches on free surface induced by geometric inaccuracy were the major reason to the failure of struts. A close-up of the as-printed and annealed (Fig. 5c and d) fracture surfaces showed that fine microstructure features in the as-printed struts while the annealed struts contained coarser features.

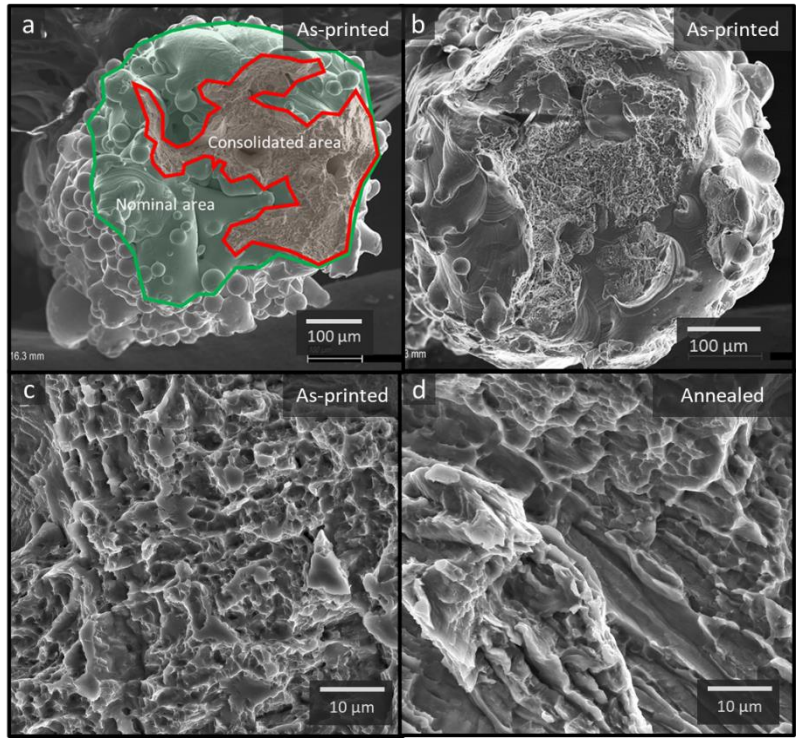

*Figure 5 – (a)(b) Fracture surfaces of two different struts showing low consolidation areas and close-ups of the fracture surfaces of the (c) as-printed and (d) annealed struts*

*3.2    Microstructure*

Microstructure was revealed by chemical etching of sectioned struts of as-printed and heat-treated meta-crystals (Fig. 6a and b). The as-printed microstructure of Ti6Al4V comprises of very fine acicular martensites (Fig. 6a), consistent with that reported in literature (Thijs et al., 2010). The $\alpha'$ martensites are responsible for the brittleness of the as-print Ti6Al4V (consistent with fine features on the fracture surface of strut shown in Fig. 5c). In comparison, the heat-treated microstructure is much coarser, note the differences in scale (Fig. 6a and c versus b and d). The annealing heat-treatment dissolved acicular $\alpha'$ martensites and transform them to β during heating to above the transus temperature. Upon slow cooling in the vacuum furnace, β transformed to larger $\alpha$ lamella colonies (Fig. 6b and d). The resulting microstructure can help to regain the ductility at the expense of strength. Both EBSD maps at the observed magnifications showed no apparent crystallographic texture.

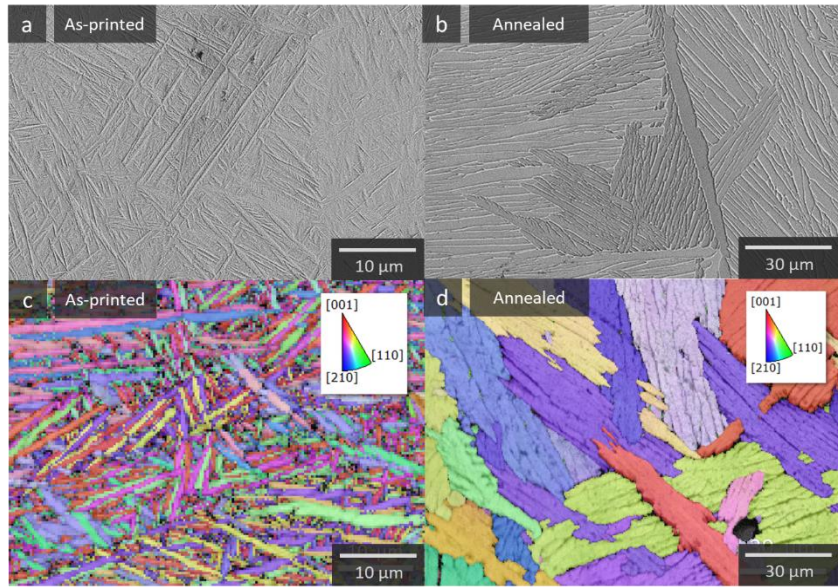

*Figure 6 - SEM micrographs and EBSD maps (IPF-build direction/out of page) showing the etched surface of the (a)(c) as-printed and (b)(d) heat-treated meta-crystals. Nb. the SEM micrographs and EBSD maps are not of the same region.*

*3.3    Mechanical properties*

The stress-strain curves of the as-printed titanium meta-crystals with 0.5 mm diameter struts are shown in Figure 7a (stress was normalised by the relative density). The measurement of toughness (Fig. 7b) shows that there is an improvement in toughness with increasing the number of meta-grains (i.e. reducing the size of meta-grains) in agreement with a previous study on meta-crystals fabricated by a PLA (Pham et al., 2019). Nevertheless, the improvement was insignificant and much less in comparison to PLA meta-crystals. All meta-crystals still exhibited classic stretching-dominated lattice mechanical behaviour: sharp post-yield stress drop and subsequent unstable response thereafter. Increasing the number of meta-grains did not noticeably improve the strength, nor the post-yield behaviour. This lack of significant improvement was explained by the DIC of the deformation (Fig. 8), which shows the deformation under increasing compressive strain in the meta-crystals containing 16 meta-grains. The analysis of the as-printed meta-crystals with 0.5 mm diameter strut revealed that the meta-grain boundaries did not arrest or deflect the initial localised deformation band, suggesting the stress concentration induced by deformation band was stronger than the strength of meta-grain boundaries and, therefore, the boundaries failed to improve the initial collapse behaviour of the meta-crystal (strength, stress drop magnitude). The deformation bands also formed on arbitrary planes and orientations to that predicted previously in (Liu et al., 2020). The unstable response and little hardening in the post-yield behaviour was attributed to the successive crumbling of lattice planes (Fig. 8).

Section 3.1 shows that increasing the strut diameter can lead to higher strut quality (Figs. 3 and 4). Figures 9a and 9b indeed shows that there is a clear trend where the strength and toughness improved with a reduction in the size of meta-grains for the meta-crystals with the 0.8 mm strut diameter, i.e. the contribution of meta-grain boundary strengthening was enhanced. The deformation of the meta-crystals with the 0.8 mm diameter struts showed formations of secondary deformation bands following the initial collapse despite the ill-defined (formation on arbitrary locations) initial deformation bands (Fig 8).

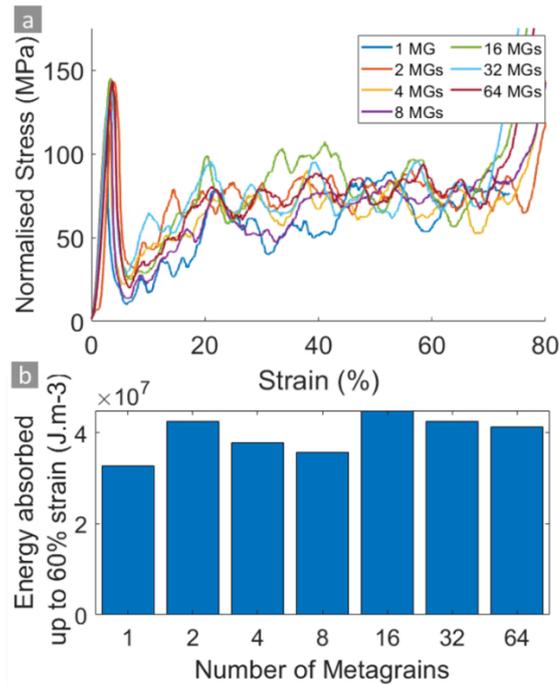

*Figure 7 - Mechanical behaviour of as-printed Ti6Al4V meta-crystals with 0.5 mm diameter. (a) The normalised stress-strain curves, (b) the total energy absorbed up to 60% strain.*

As mentioned, the deformation behaviour of the meta-crystals design was governed by the behaviour of the base material. Ti6Al4V fabricated by powder bed fusion is known to be very brittle (Xu et al., 2015). The rapid cooling in L-PBF resulted in $\alpha'$ martensite microstructure (Fig. 6a). Together with the high thermal gradient (estimated up to $10^7$ K/m) in L-PBF, fine solidification microstructure containing dense dislocations is often formed in as-printed alloys (Pham et al., 2017). Such microstructure caused high strength, but low ductility (Harun et al., 2018). Together with the predominant notch-like defects, the brittleness of fine struts caused successive lattice planes perpendicular to the loading direction crumbled with increasing strain (Fig. 8). Formation of localised deformation bands in arbitrary orientations and crumbling of lattice planes suggest that the individual struts were very weak, incapable of transferring the load to the bulk of the structure and would catastrophically fracture.

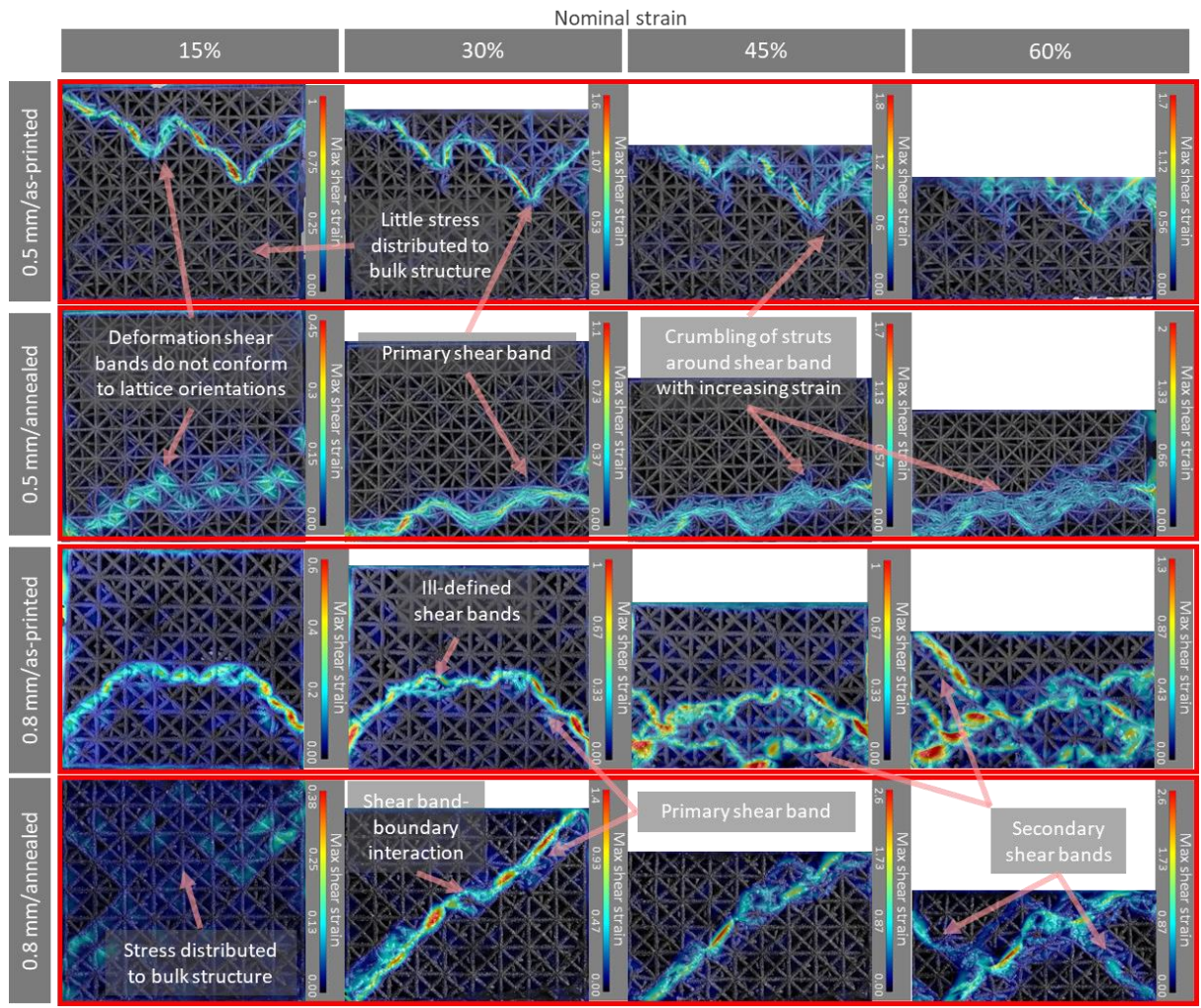

*Figure 8 - DIC showing the deformation of the 16 meta-grains meta-crystals under increasing compressive strain. From left to right of the same row: Deformation fields at 15 %, 30 %, 45 % and 60 %. First and second rows: the meta-crystal with the 0.5 mm strut diameter of the as-printed and annealed conditions, respectively. Third and fourth rows: the meta-crystal with the 0.8 mm strut diameter of the as-printed and annealed conditions, respectively*

Heat treatment was done to regain the ductility and verify if the meta-grain boundary strengthening was enhanced. Annealing heat-treatment significantly reduced the yield strength, suggesting the residual stress was successfully removed after annealing thanks to the change in microstructure. In addition, the relative magnitude of the initial stress drop of all meta-crystals lessened and their post-yield behaviour became less erratic compared to the as-printed meta-crystals (Fig.9a). The annealing also enhanced the contribution of meta-grains with more noticeably increased toughness from meta-crystals containing more meta-grains (Fig 9c). However, deformation of the annealed meta-crystals with 0.5 mm diameter struts still showed crumbling of struts after the formation of the initial localised bands (Fig 8, second row), indicating the process defect is the predominant factor governing the behaviour of the meta-crystals.

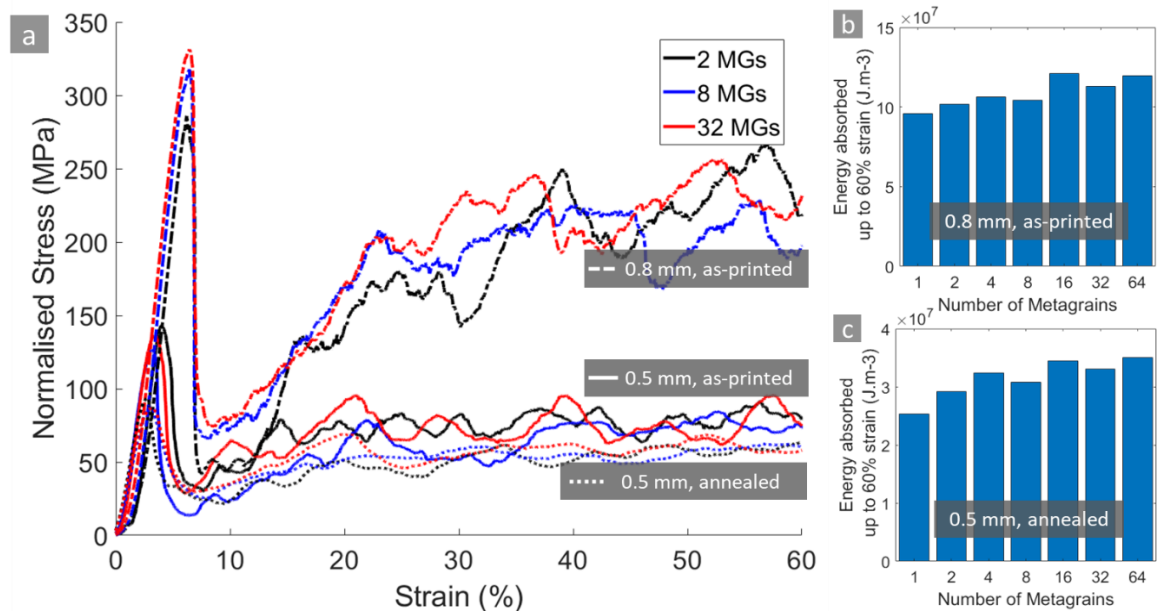

Figure 9 - (a) Normalised compressive stress-strain curves of 2, 8 and 32 MGs representative meta-crystals: 0.5 mm/as-printed (solid), 0.5 mm/annealed (dotted), and 0.8 mm/as-printed (dashed), and the total energy absorbed up to 60% strain of the meta-crystals with (b) 0.8 mm diameter, as-printed and (c) the 0.5 mm diameter, annealed.

The synergistic effect of both increasing the strut diameter and heat-treatment was demonstrated when studying the behaviour of the annealed and as-printed meta-crystals with the strut diameter of 0.8 mm (Fig. 10a). Despite the heat-treatment, there is no reduction in the peak strength as observed for the annealed meta-crystals with the strut diameter of 0.5 mm. However, the benefit of the heat-treatment remained -which shows a clear reduction in the initial stress drop in the stress-strain response (Fig 10a). The significant decrease in the initial post-yield collapse was seen with more meta-grain, suggesting the meta-grain boundaries played more influential for the annealed meta-crystals. In particular, the meta-crystals with 32 meta-grains behaved much more stable, further confirming the role of meta-grain boundaries once the process defects were reduced and the ductility of the base material was regained. DIC shows the formation of a well-defined initial localised band followed by secondary localised bands with increasing nominal strains which helps to re-distribute the deformation to the bulk structure (Fig 8, fourth row).

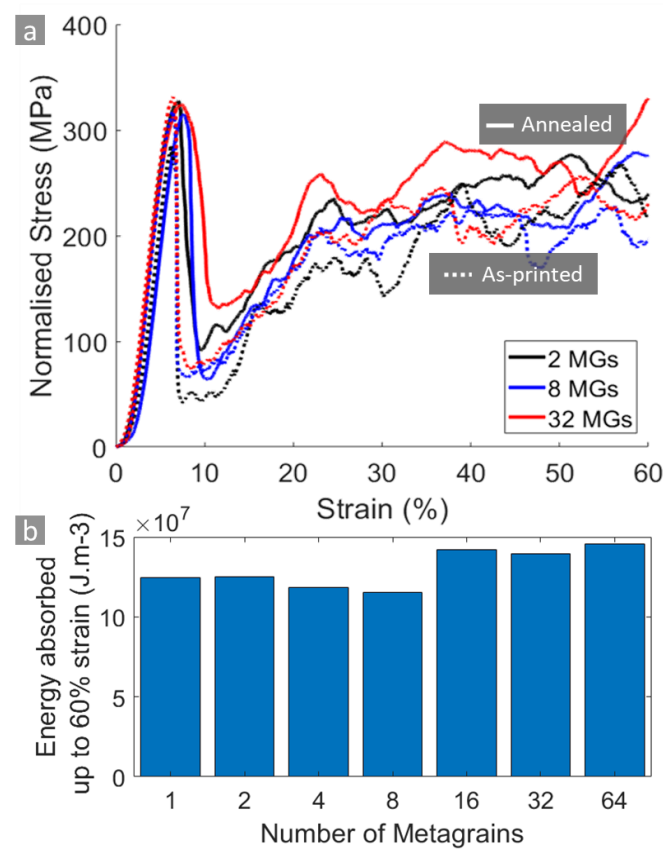

*Figure 10 - (a) Normalised compressive stress-strain curve of the annealed (solid) and the as-printed (dotted) 2, 8, and 32 MGs meta-crystals with 0.8 mm strut diameter. (b) the total energy absorbed up to 60% strain of the annealed meta-crystals with 0.8 mm strut diameter.*

4   Discussions

*4.1   Effect of process defects on the boundary strengthening in Ti6Al4V meta-crystals*

A previous study showed that the mimicry of polygrain microstructure is effective in strengthening architected materials and eliminating the post-yielding collapse (Fig. 1b) (Pham et al., 2019). The addition of meta-grain boundaries was able to deflect and scatter localised deformation bands, substantially increasing the strength and the damage tolerance of architected materials made from polylactic acid (PLA). However, when such meta-crystals are fabricated from Ti6Al4V by LPBF, the effectiveness of the meta-grain boundaries significantly reduced. The performance of all meta-crystals worsened including the single meta-grain meta-crystal, which did not contain strengthening boundaries, implying that the differences in the material and the processing were responsible for the degraded performance. PLA struts can be fabricated with high geometric accuracy and high consolidation, helping to eliminate the adverse effect induced by processing defects and enable an influential role of meta-grain boundaries in improving the strength and post-yield stability of meta-crystals. In contrast, Ti6Al4V struts were fabricated with low accuracy, Figures 3 and 4 (in particular insufficient fusion occurred at free surface, causing high stress-raiser notches (Fig. 3 and 4a, b). The most damaging defects can be seen at the fracture surface, which shows poor consolidation giving rise to notch-like asperities on the free surface. Such abundant notch-like defects resulted in the loss in load-bearing areas, causing overload and fast fracture during loading for the consolidated remain ligaments, preventing the strengthening induced by the meta-grain boundaries. The DIC of the compression (Fig. 8) shows that no defined shear bands formed in the as-printed meta-crystals with 0.5 mm diameter struts, unlike in PLA meta-crystals. Instead, lattice struts crumbled in a layer-by-layer manner successively, confirming that the struts were weak and unable to transfer the load to the remainder of the structure.

It was important to reduce the significance of such defects to ensure the efficacy of the meta-crystal designs. The increase the strut diameter was shown to help achieving better quality of struts (Figs. 3, 4a, b and 9a). The increase in the strut diameter can also mitigate the adverse effect of defects via reducing the dimensions of notches in relation to the that of consolidated ligaments. Increasing the strut diameter ($d$) while the lattice spacing is kept constant leads to a smaller slenderness ratio - $l/r$ where $l$ is the length of strut and $r$ is the radius of gyration ($r$

= *d*/4). A slenderer strut is more likely to buckle, as given by Euler buckling (Eq. 3) (where k²L² is the effective length, E is the elastic modulus, and I is the second moment of area).

$$P_{cr} = \frac{\pi^2 \mathrm{EI}}{k^2 L^2} \tag{3}$$

where, $I = \frac{\pi d^4}{64}$

The normalised stress (using the average measured $\bar{\rho}$) can be expressed as follows:

$$\sigma_{norm} = \frac{P_{cr}}{A\bar{\rho}} = \frac{\pi^2 \mathrm{E}(\pi d^4/64)/(k^2 L^2)}{(\pi d^2/4)\cdot\bar{\rho}} = \frac{\pi^2 \mathrm{E} d^2}{16 k^2 L^2 \bar{\rho}} = C \cdot \frac{d^2}{\bar{\rho}} \tag{4}$$

where, $C = \frac{\pi^2 \mathrm{E}}{16 k^2 L^2}$

The normalised stress required for buckling was calculated to be approximately very similar (2.67C and 2.66C) for the 0.5 mm and the 0.8 mm strut, respectively. Therefore, the drastic increase in peak stress with an increase in strut diameter seen in Figure 9a can be mostly attributed to the better strut quality and a reduced role of notch-like asperities (Fig. 3 and 4), i.e. lowering the contribution of the process defects. The better quality with the larger strut diameter effectively avoided premature fracture of struts. While there the response remained erratic, an overall hardening behaviour was observed (Fig. 9a). This was attributed to the formation of secondary deformation bands with increasing strain as seen in the DIC analysis (Fig. 8, third row). By avoiding early fracture of strut, the deformation was homogenised by utilising the entire architecture, helping to increase the impact of the meta-crystal design. As a result of the increasingly influential role of the meta-grain boundaries, there is clear improvement in the mechanical behaviour of the meta-crystals with 0.8 mm (Fig. 11b) and the energy absorption behaviour with increasing number of meta-grains (Fig. 9a, b, c).

*4.2    Effect of the crystalline microstructure*

As discuss in Section 4.1, the detrimental effects of defects were also amplified by the brittleness of the as-printed Ti6Al4V. The brittleness of the strut was attributed to the presence of acicular α' martensite seen in the EBSD map (Fig. 6c) and significant internal stresses built up during rapid cooling with large thermal gradient in L-PBF (Harun et al., 2018; Sames et al., 2016). To counteract this, the meta-crystals were subjected to an annealing heat-treatment which resulted in a coarse (α+β) microstructure and an increased ductility (Fig. 6b). The heat-treatment also relieved the internal stresses built up during processing, resulting in the reduction in the yield strength (Fig 9a). Both the increased ductility and diminished internal stresses should minimize the tendency for the struts to fast fracture at low strain. The deformation behaviour of the annealed meta-crystals with the 0.5 mm strut diameter showed that this was not the case. While the magnitude of the collapse reduced thanks to the decreased yield strength, the meta-crystals still crumbled successively in a layer-by-layer manner (Fig 9a), suggesting that the process defects were too substantial to be mitigated by improving the microstructure. In addition, annealing alone is not capable of removing the process defects (Fig. 3). As seen previously, the defects caused early fracture of the struts and compromised the effects of the meta-grain boundaries. Although a small improvement was seen in the toughness with an increasing number of meta-grains, the change in microstructure by heat-treatment was insufficient for the given as-print quality of the meta-crystals. This is supported by the fracture behaviour seen on the fracture surface of the annealed strut (Fig. 5d). A previous study shows that coarse lamellar (α+β) only increased the elongation to failure by ~6% compared to acicular α' martensite microstructure (Xu et al., 2017). Even with the heat-treatment and increased ductility, the (α+β) titanium is still rather brittle compared to other materials such as 316L stainless steel that is ductile even in as-printed condition (Pham et al., 2017) or aluminium alloys (Croteau et al., 2018). The benefits of the heat-treatment are apparently limited to reducing the severity of the stress drops thanks to lowered yield strength and hence a less serrated post-yield stress response.

*4.3 Enabling an influential role of meta-grain boundaries strengthening in the mechanical behaviour of meta-crystals*

It has been clear that it is necessary to fabricate quality lattice struts to realise the potential of the crystal-inspired approach in developing high damage-tolerant metallic meta-crystals. It is not practical to use hot isostatic pressing (HIP) to close pores for lattice materials. In addition, the most severe defects were found were notch-like defects on free surface which was known not to be removed by HIP (Tammas-Williams et al., 2016). The most straightforward solution is to optimise the process parameters to minimise the defects. Nevertheless, it would remain difficult to eliminate them entirely (particularly for intricate and fine structures). The tendency for defect formation is affected by other factors such as the part geometry or the choice of alloy. Figures 3 and 4 show that a better choice of the strut dimensions can be effective in achieving good quality lattice struts, resulting in improving the mechanical performance of meta-crystals and enhancing the contribution of meta-grain boundaries. However, the increase in the strut diameter can result in an increase in the relative density of meta-crystals. Regarding the choice of alloy, some materials were shown to be more 'printable' (Piglione et al., 2018) and have good ductility despite numerous pores in the as-print condition (Qiu et al., 2018). Similar to the geometric design, the right choice of material can help to realise the full potential of meta-crystal architecture. For a high strength, but low ductile material like Ti6Al4V, the crystal-like architecture will need to be modified to accommodate the material's behaviour. The boundaries between meta-grains in the current design creates a twin boundary between meta-grains. DIC results in Figure 8 shows that the deformation band can easily penetrates the twin boundary with no change to the direction due to the symmetrical nature of lattice across the twin boundary. It is known in physical metallurgy that the slip transmission across a grain boundary relates to the misorientation and coherency of the boundary (Brandon, 1969; Brandon et al., 1964). Changing the misorientation alters the boundary coherency. As such, engineering the misorientation between meta-grains can significantly increase the boundary hardening effect. Alternatively, the boundary hardening effect via arresting and deflecting the propagation of deformation bands can be increased by strengthening the boundary by means such as increasing the strut diameter or reducing the lattice spacing of the boundary.

## 5    Conclusions

This study investigated the effect of the multiscale hierarchical lattice structure from the polycrystalline microstructure, architected polygrain-like mesostructures and process defects on the mechanical behaviour of Ti6Al4V meta-crystals fabricated via laser powder bed fusion. Meta-grain boundaries were shown to significantly improve the strength of PLA meta-crystals. However, the architectural benefits were diminished in Ti6Al4V meta-crystals due to the predominance of process defects and the brittleness of the as-printed material. In particular, lack-of-fusion on free surface caused notch-like defects that were found to be most influential. Such defects led to stress concentration and significantly reduced the load bearing areas, causing overload and fast fracture of consolidated ligaments, drastically weakening the struts. Increasing the strut diameter effectively enhanced the quality of as-printed struts and reduced the roles of defects, enabling more contribution from meta-grain boundaries to improving the mechanical performance of the meta-crystals. In addition, the brittleness of Ti6Al4V meta-crystals was related to the as-printed microstructure (in particular acicular $\alpha'$ martensite) and high internal stresses (due to large thermal gradient in L-PBF). The annealing heat-treatment increased the ductility and reduced the internal stresses thanks to the transformation of $\alpha'$ martensite to ($\alpha+\beta$) microstructure. Although such enhancement in microstructure was not sufficient for the thin strut diameter (0.5mm), the enhancements were noticeable in the post-yield performance of the meta-crystals for the thicker struts (0.8mm). Most importantly, both the heat-treatment and the strut thickening synergistically increased the effectiveness of the meta-grain boundaries. The meta-grain boundaries significantly improved the toughness and post-yield behaviour of annealed meta-crystals with the 0.8 mm diameter struts. The post-yield stress drop was markedly reduced, and the Ti6Al4V meta-crystals were much more stable after yield with noticeable hardening behaviour seen thereafter. The results showed that the right choice of geometry and crystalline microstructure can enable the use of the crystal-inspired approach to create damage-tolerant meta-materials with high mechanical performance.

## Acknowledgements

This work is supported by Imperial College London's The President's Excellence Fund for Frontier Research.

The authors would like to acknowledge Betatype Ltd. for the fabrication of the samples.